# Raman spectroscopy and *In-situ* Raman spectroelectrochemistry of isotopically engineered graphene systems

*Otakar Frank[1], Mildred S. Dresselhaus[2,3], Martin Kalbac[1*]*

[1] J. Heyrovský Institute of Physical Chemistry of the AS CR, v.v.i., Dolejškova 3, CZ-18223 Prague 8, Czech Republic

[2] Department of Electrical Engineering and Computer Science, MIT, Cambridge, Massachusetts 02139, USA

[3] Department of Physics, MIT, Cambridge, Massachusetts, 02139, USA

*Corresponding author: Tel: 420 2 6605 3804; Fax: 420 2 8658 2307; E-mail: kalbac@jh-inst.cas.cz

**Conspectus**

The unique properties of graphene offer immense opportunities for applications to many scientific fields, as well as societal needs, beyond our present imagination. One of the important features of graphene is the relatively simple tunability of its electronic structure, an asset which extends the usability of graphene even further beyond present experience. A direct injection of charge carriers into the conduction or valence bands, *i.e.*, doping, represents a viable way of shifting the Fermi level. In particular, the electrochemical doping should be the method of choice, when higher doping levels are desired and when a firm control of experimental conditions is needed.

In this Account, we focus on the electrochemistry of graphene in combination with *in-situ* Raman spectroscopy, *i.e.*, the *in-situ* Raman spectroelectrochemistry. Such a combination of methods is indeed very powerful, since Raman spectroscopy can readily monitor not only the changes in the doping level, but it can give information also on eventual stress or disorder in the material. However, when employing Raman spectroscopy, one of its main strengths lies in the utilization of isotope engineering during the chemical vapor deposition (CVD) growth of the graphene samples. The *in-situ* Raman spectroelectrochemical study of multi-layered systems with smartly designed isotope compositions in individual layers can provide a plethora of knowledge about the mutual interactions: (i) between the graphene layers themselves, (ii) between graphene layers and their directly adjacent environment (*e.g.*, substrate or electrolyte), and (iii) between graphene layers and their extended environment, which is separated from the layer by a certain number of additional graphene layers. In this Account, we show a few examples of such studies, from monolayer to two-layer and three-layer specimens, and considering both turbostratic and AB interlayer ordering. Furthermore, the concept and the method can be extended further beyond the 3-layer systems, as, for example, to heterostructures containing other 2-D materials beyond graphene.

In spite of a great deal of important results has been unraveled so far through the *in-situ* spectroelectrochemistry of graphene based systems, there still lie many intriguing challenges immediately ahead. For example, apart from the aforementioned 2-D heterostructures, a substantial effort should be put into a more detailed exploration of misoriented (twisted) bilayer or trilayer graphenes. Marching from the oriented, AB-stacked to AA-stacked, bilayers, every single angular increment of the twist between the layers creates a new system in terms of its electronic properties. Mapping those properties and interlayer interactions as a function of the twist angle represents a sizeable task, yet the reward might be the path towards the realization of various types of advanced devices.

And last but not least, understanding the electrochemistry of graphene paves the way towards a controlled and targeted functionalization of graphene through redox reactions, especially when equipped with the possibility of an instantaneous monitoring of the thus introduced changes to the electronic structure of the system.

**Introduction**

Graphene has attracted attention due to its unique properties that are encouraging for numerous applications in nanoelectronics. However, the up-and-coming utilization of graphene necessitates a deeper comprehension of its electronic properties, both in its neutral and charged states. One of the great advantages of graphene is the tunability of its optical and transport properties by doping, which leads to a shift of the Fermi level, by adding carriers to either the conduction or valence bands. The ability to shift the Fermi level presents a vital degree of freedom which augments the range of potential applications of this unique and exceptional material.

In general, graphene can be doped chemically (also called molecular doping),[1] electrochemically,[2-5] by electrostatic gating,[6,7] or by a direct introduction of heteroatoms into the lattice, see Figure 1.[8] Till now, such experiments on graphene have usually been performed using electrochemical or electrostatic doping since these methods provide a direct approach to manipulate the Fermi level of graphene.[2-5,7,9-11] Electrostatic backgating[7,12] is used for its simplicity and a straight application channel en route to field-effect transistors,[13] but it has several downsides. Firstly, electrostatic gating relies on the properties of the dielectric layer. Because the doping efficiency is commonly modest and therefore a high voltage needs to be applied, the accessible magnitude of doping levels is restricted. For example, Yan et al.[7] used a gate voltage range of -80 to 80 V to reach carrier densities from 8 to $-4\times10^{12}$ cm$^{-1}$. Additionally, a high applied voltage can cause a charge trapping by the substrate, thereby altering its properties. The experimental results are then more challenging to interpret and to reproduce from one sample to another. In contrast, electrochemical doping is efficient, in that an electrode potential of ±1.5 V suffices for typical experiments, thereby reaching carrier concentrations of ~±5×10$^{13}$ cm$^{-2}$.[3,5,9] However, further extension of the doping range is highly desirable and can be achieved using liquid electrolyte in a conjunction with a protective layer.[14,15] In a standard electrochemical experiment, the complications with charge trapping are mitigated, because the carriers are transferred to the sample through an ohmic contact and recompensed by an electrolyte counterion. Morever, graphene responds to electrochemical doping quickly enough, and thus the measurement speed is generally hampered only by the time required to get a sufficient intensity of the signal in the spectrum. On the other hand, the electrochemical method demands distinct cell geometry, high chemical purity of the electrolyte salt and solvent, and high quality electrodes.[16] Furthermore, a three electrode set-up with a reference electrode (apart from the counter and working electrodes) has to be utilized to precisely control the applied voltage.[17]

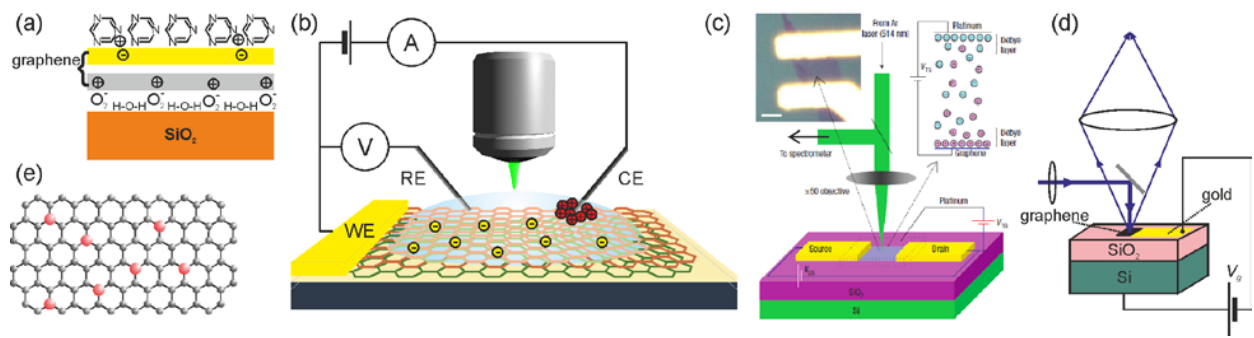

**Figure 1.** Sketches of different doping experiments conducted on graphene: (a) chemical doping, adapted with permission from Ref [1], (b) electrochemical 3-electrode setup used, *e.g.*, in Refs [3,5], (c) electrochemical gating setup, reprinted with permission from Ref. [2], (d) electrostatic gating setup, adapted with permission from Ref. [7], (e) incorporation of atoms into the graphene lattice.

Before the rise of graphene,[18] the *in-situ* combination of Raman spectroscopy with electrochemistry, *i.e.*, Raman spectroelectrochemistry, had already proven useful for the study of fullerenes and carbon nanotubes.[16] The important features monitored in the Raman spectra of pristine graphene (Figure 2) are the symmetry-allowed G and G' modes[19,20] (the latter also termed as the 2D mode[19]). They can be found in the Raman spectra of all graphene-derived materials; however, their particular Raman shifts, line-widths and intensities are affected by the laser excitation energy, number of graphene layers, doping, strain, *etc*.[6,21] The D line can also appear in Raman spectra of some graphene samples indicating the presence of symmetry-breaking perturbations. A few recent studies describe the spectroelectrochemical behavior of the D band in graphene,[22-24] showing the tunability of the D peak presence (both reversibly and irreversibly),[23,24] but also its response to the applied potential.[22]

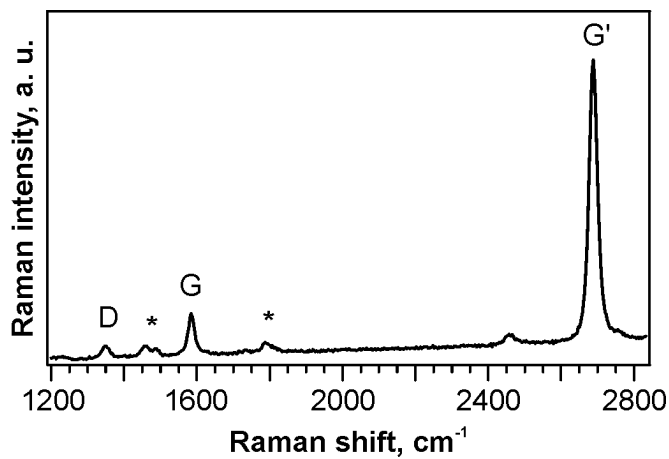

**Figure 2.** Raman spectrum of single-layer CVD graphene at 0 V, excited by 2.33 eV laser excitation energy, in an electrochemical environment. Asterisks indicate Raman bands of the electrolyte. Reprinted with permission from Ref [5]. Copyright 2010 American Chemical Society.

For fundamental research, mechanically cleaved graphene[18] has usually been favored over CVD graphene[25] mostly owing to the lower quality of CVD graphene in the early stages of its development. However, with the recent rapid progress in both CVD growth (graphene domains with sizes over 1mm$^2$ now available[26]), and the consecutive graphene transfer, the quality contrast between the samples have become smaller. On top of that, CVD growth provides one special tool unaccessible by mechanical cleavage – namely that of carbon isotope labeling,[27] which is the focus of this review. Thus the CVD-prepared graphene significantly simplifies the sample processing, thereby facilitating more detailed studies with such samples.

Since CVD graphene is of high current interest, we discuss in this review the electrochemical results obtained thus far when using CVD graphene. Although these results are similar to those obtained on cleaved graphene samples, one should expect some minor differences due to the grain boundaries, and wrinkles or folds that are more common in CVD graphene. This review will successively describe the cases of monolayer graphene (referred to as 1-LG), sequentially transferred two- and three-layer graphene (2-LG and 3-LG, respectively), and bilayer graphene grown in one step (AB 2-LG or turbostratic 2-LG, depending on the relative crystallographic orientation between the two layers).

**Monolayer graphene**

In general, the G band frequency shift in doped graphene is governed by the alterations of the C-C bond strength and by the phonon energy renormalization.[28] In graphene, the similarity of timescales of electron and phonon dynamics allows a coupling between lattice vibrations and Dirac fermions. Therefore the description of the G band phonons by the adiabatic Born-Oppenheimer approximation is not successful,[19,28] and the time-dependent perturbation theory is used instead. In this description, an electron is first excited from the valence band (referred to as π) to a conduction band (π*) by absorbing a phonon, and an electron-hole pair is thus created. The electron and the hole then recombine and emit a phonon, whose lifetime and frequency are now notably altered by this second-order process.[6] As a consequence, the energies of both the phonons and the carriers are renormalized. In doped graphene, the creation of electron-hole pairs can be quenched, as the Fermi energy $E_F$ is shifted from the Dirac point.[28] The change of the G band frequency should be the same for positive and negative doping because of electron-hole symmetry with respect to the Dirac point. However, the C-C bond strength is also influenced by doping.[7] Electrons are removed from antibonding orbitals upon positive doping, which results in a hardening of the G mode. Conversely, electrons are added to the antibonding orbitals upon negative doping, and therefore a softening of the Raman G mode frequency ($\omega_G$) is expected, as is known from the studies of graphite intercalation compounds.[29]

In graphene, the effects of both renormalization of the phonon energy and change of the C-C bond strength are superimposed in the experimental data. For positive charging, the two effects add up in an upshift of $\omega_G$.[2,5] In contrast, the effects have an opposite sign of the G band frequency shift during negative charging.[2,5] These arguments are in line with the electrochemical experiments, where $\omega_G$ was found to rise monotonically at positive electrode potentials and non-monotonically at negative potentials.[2] Additionally, the measured shift never becomes as high for the negative doping as it does for the positive doping.[5]

As mentioned above, the evolution of the Raman G and G' modes upon electrochemical doping has been firstly examined for mechanically cleaved graphene samples.[2] The obtained data are fully comparable to the results gathered for the CVD graphene,[5] verifying that the electronic structure of graphene is tunable independently of its preparation method. Nevertheless, it should be noted that different authors may use different electrolytes and also different electrochemical set-ups, and these differences explain some of the inconsistencies between the published results, which were observed by different research groups. More specifically, the doping efficiency can often be such an issue. Hence, much higher electrode potentials were needed in the case of less efficient electrolytes[2] to achieve the same effect as in the case of highly efficient electrolytes.[14]

In spectroelectrochemistry measurements,[16] when a high electrochemical potential is needed, non-aqueous electrolytes are preferred over the aqueous ones due to an early onset of water decomposition both at positive and negative potentials, giving limited potential window of less than 2V for the electrochemical study of turbostratic $sp^2$ materials.[24] If allowed by the sample conditions, *i.e.*, by a good adhesion to the substrate, a liquid electrolyte is advantageous due to a generally better conductivity compared to polymer-based electrolytes.

An example ca be given for electron doping, where, in Ref [5], the electrode potential of -1.2 V caused the maximum Raman G band shift of 1604 cm$^{-1}$ (Figure 3, left), whereas an electrode potential of about -4 V was needed to attain the same $\omega_G$ in an earlier work.[2] It has to be mentioned that common electrolytes start to undergo irreversible changes at potentials much lower than -4 V. Consequently, this means that measurements conducted at such potentials were probably not under an equilibrium state and the polarization of the graphene working electrode was not perfectly quantitative from an electrochemical standpoint. Obtaining a high doping efficiency is thus essential for proper analysis of the electrochemical data collected at larger doping levels, because electrolyte or electrode instabilities may take place at high potentials and they need to be avoided for obtaining reproducible results.

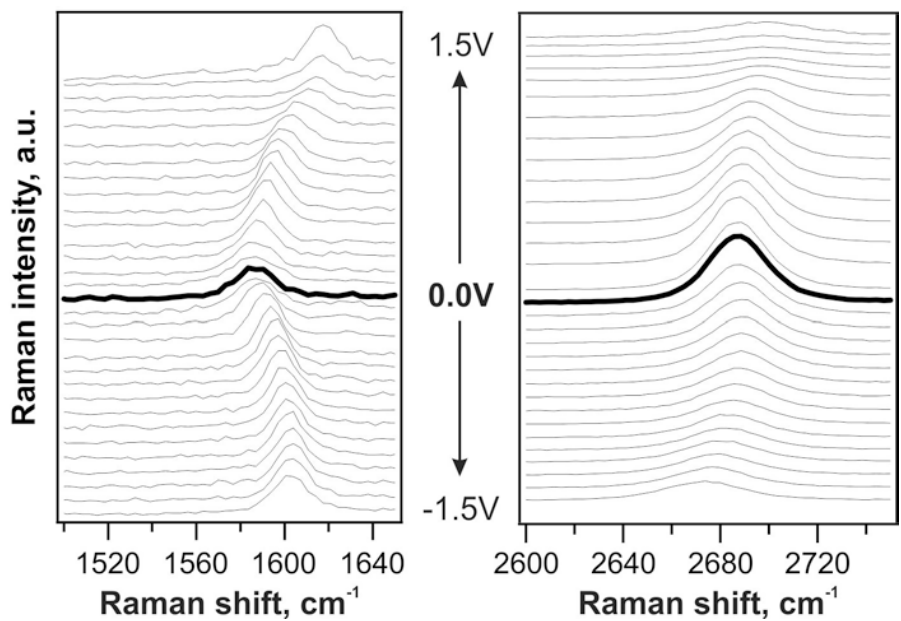

**Figure 3.** *In-situ* Raman spectroelectrochemical data for $\omega_G$ (left) and $\omega_{G'}$ (right) on 1-LG, excited by 2.33 eV laser radiation. The separation between traces is 0.1 V. Adapted with permission from Ref [5]. Copyright 2010 American Chemical Society.

We also note that throughout this text, the values of the potential refer to the working (graphene) electrode. Therefore hole doping is denoted by positive potential values and electron doping by negative values. Such code is customary in electrochemical works.[3-5,16] That is in contrast to some works, where the potentials values refer to the reference (gate) electrode.[2,10,11] Such notation gives the opposite sign to the potential values, compared to electrochemistry. Hence, in this other case, hole doping of graphene is denoted by negative potential values, while electron doping of graphene is denoted by positive electrode potentials values. Thus, when considering electrochemical measurements of graphene, it is important for authors to define the signs corresponding to electron and hole doping, and for the reader to be aware of possible differences in notations between one report and another.

The Raman G' mode frequency ($\omega_{G'}$) reacts sensitively to doping as well, however, differently than the G mode (Fig. 3, right). Increasing magnitude of the positive potentials results in an increase of $\omega_{G'}$, while at negative potentials $\omega_{G'}$ increases slightly at first, followed by a rather large downshift. In the range of 0 to 1 V, the potential-dependent change of the G' mode frequency exhibits a slope of $\Delta\omega_{G'}/\Delta V = 9$ cm$^{-1}$/V, while the corresponding slope of the G mode in the same potential range was $\Delta\omega_G/\Delta V = 18$ cm$^{-1}$/V.[5] The ratio between $\Delta\omega_{G'}/\Delta V$ and $\Delta\omega_G/\Delta V$ gives 0.5, which corresponds very well with theoretical prediction.[30] The change of $\omega_{G'}$ upon doping comprises the effects of variations in the C-C bond strength, the electron-phonon coupling and electron-electron interactions. It should be noted that the distinct mutual changes of $\Delta\omega_G$ vs $\Delta\omega_{G'}$ induced by doping and strain can be used to disentangle these two effects in graphene under various conditions, such as the conditions caused by (i) varying the interaction between CVD graphene and underlying copper single crystals,[31] (ii) modifying the interface between exfoliated graphene and the Si/SiO$_2$ substrate,[32] or (iii) by back-gating epitaxial graphene on SiC.[33]

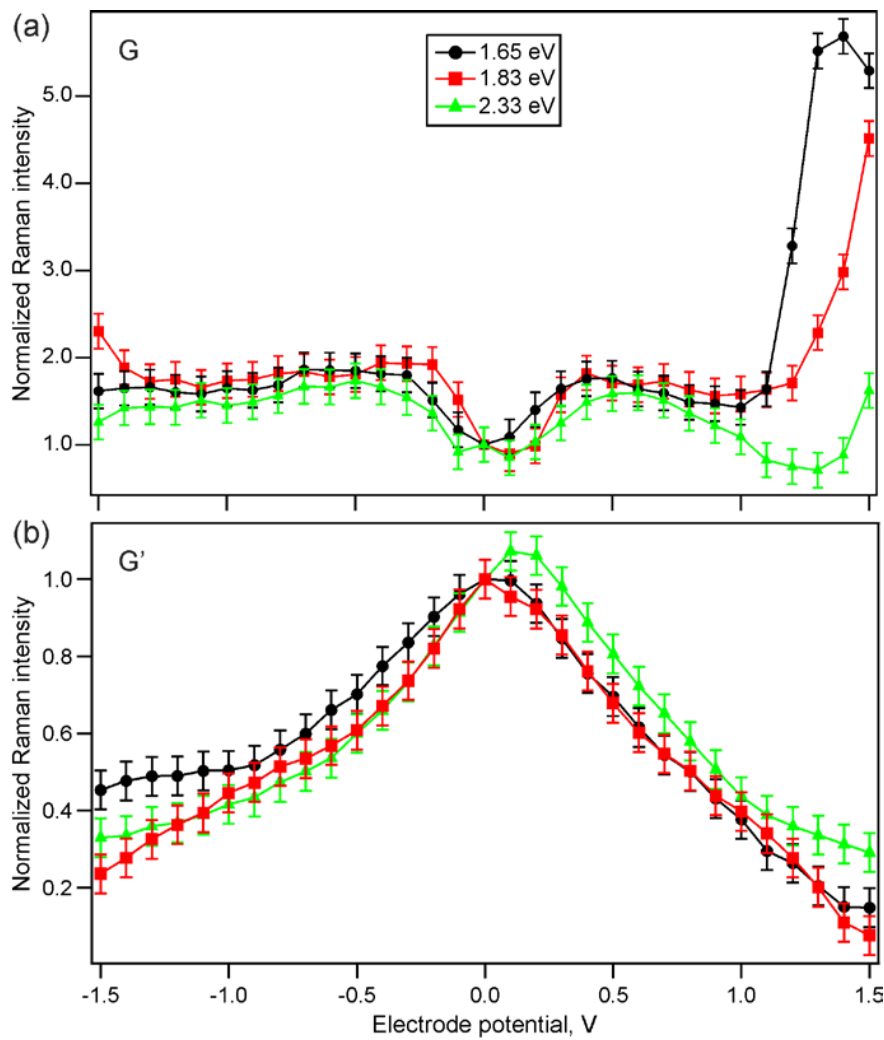

**Figure 4.** Raman intensity vs. electrode potential profiles for the (a) G and (b) G' modes at various laser excitation energies. Reprinted with permission from Ref [5]. Copyright 2010 American Chemical Society.

The intensities (areas) of both the G and G' Raman modes for graphene manifest a considerable and mode-specific evolution in response to the electrode potential. The most notable feature is an extreme enhancement of the Raman G band intensity at positive potentials over 1.0 V (Fig.4 top) that is not observed for electron doping. This intensity increase depends sensitively on the laser excitation energy, where it was strongest and with the earliest onset for the lowest excitation energy used (1.65 eV).[5]

In undoped graphene the phonons can dissipate energy through the creation of an electron-hole pair.[28,30] In charged graphene this process is quenched since the final state is either occupied (for electron doping) or empty (for hole doping), which causes the G band narrowing.[7,28,30] After the removal of Kohn anomaly,[5] the intensity of the G band does not change upon farther electron doping and up to 1.0 V for hole doping, in agreement with previous theoretical predictions and experimental results.[7] However, a similar enhancement of the signal at high positive potentials was also observed in an experiment using an ionic liquid electrolyte.[34] Such appearance of the dramatic intensity increase is in accordance with theoretical work by Basko.[35] When the Fermi level approaches $E_{laser}/2$, the matrix element for the G band should increase.[35] Additionally, the Fermi level will reach $E_{laser}/2$ at smaller values of positive potential, when smaller laser excitation energy is used. The excitation energy dependence of the enhancement observed in Fig. 4 clearly corresponds to the latter case.

The Raman G' band intensity monotonically decreases for both doping directions, yet at slightly different rates (Fig. 4 bottom).[5] Basko *et al.* proposed a proportionality between the G' band intensity and the electron/hole inelastic scattering rate.[36] As the charging increases the number of carriers, the chance of a scattering event increases too, which should cause the observed decrease of the Raman G' band intensity.

**Two and three layer graphene**

Controlling and manipulating the doping state of individual layers in few layer graphene samples belongs to the crucial tasks in the graphene research. Probing the doping state of individual layers can be realized by a new concept which combines Raman spectroscopy and isotope labeling. The isotope labeling approach allows an experimentalist to prepare individual graphene layers with a distinct content for each carbon isotope.[27] We review here the results obtained on 2-LG[3] and 3-LG[4] but the approach may be used for even higher number of graphene layers.

Figure 5 shows an example of Raman spectra measured *in-situ* on electrochemically doped 2-LG sample in which the bottom layer (in contact with $SiO_2$) was graphene containing predominantly $^{12}C$ isotope (*i.e.*, the natural isotope composition) and the top-layer contained 99% of $^{13}C$ isotope (*i.e.,* the purity of the purchased chemical). In both layers the frequencies and intensities of the Raman G and G' features vary significantly with the changes of electrode potential,[3] as can be seen in Fig. 5.

The G band is upshifted by 32-33 cm$^{-1}$ at +1.5 V both for the $^{12}C$ bottom layer and $^{13}C$ top layer, and by 17 cm$^{-1}$ at -1.5 V, again, regardless of the layer position. In the same potential range, the G' mode upshifts by 9 cm$^{-1}$ at positive potential and downshifts by 5-6 cm$^{-1}$ at negative potential, regardless of the layer. Hence, it is obvious that the evolution of both the G mode and the G' mode with applied positive and negative potentials is alike for both layers.

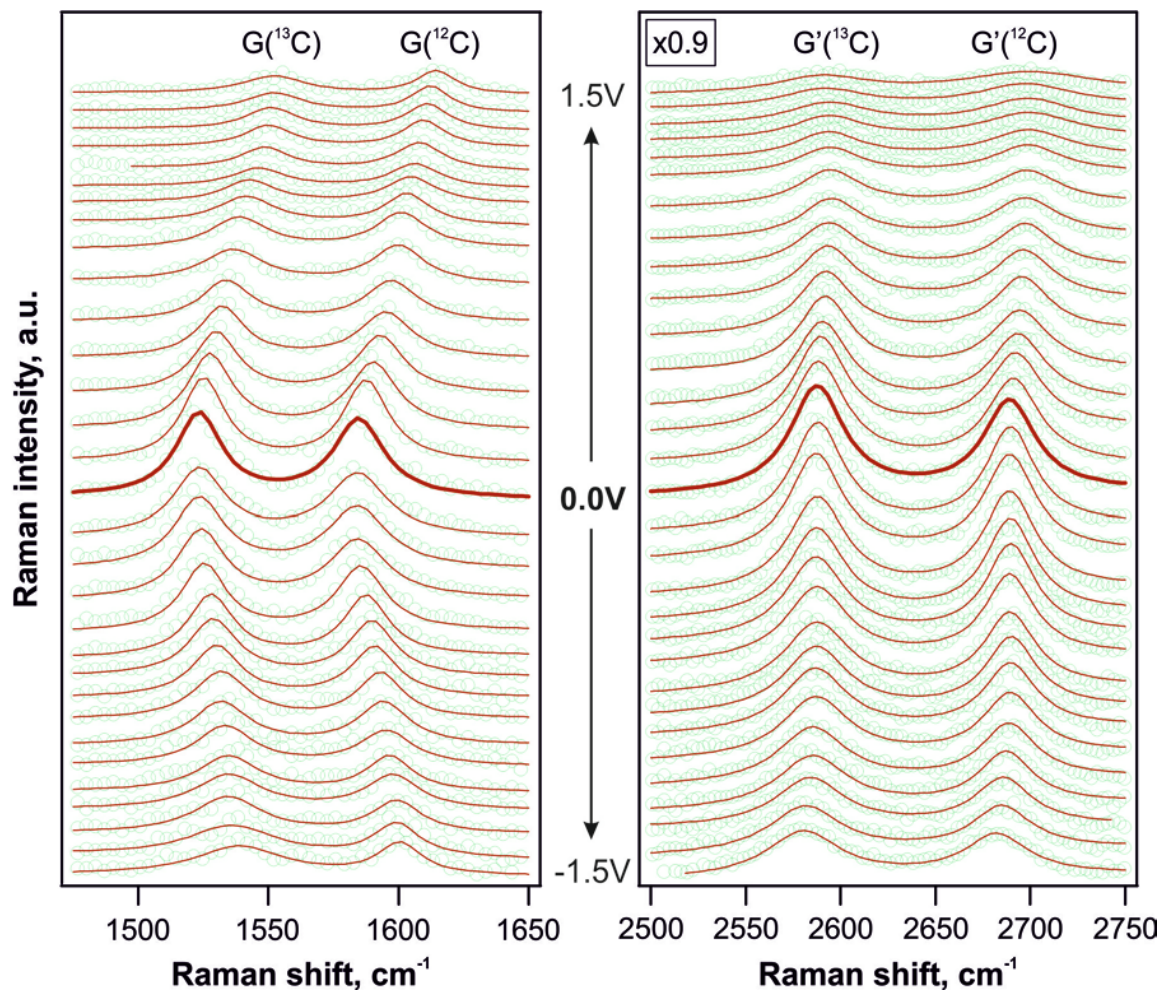

**Figure 5.** *In-situ* Raman spectroelectrochemical data for $\omega_G$ (left) and $\omega_{G'}$ (right) of a 2-LG sample, excited by 2.33 eV laser excitation energy. The circles represent the experimental points and the solid line is their fit with a Lorentzian lineshape. Adapted with permission from Reference [3]. Copyright 2011 American Chemical Society.

The same electrode potential dependence of $\omega_G$ and $\omega_{G'}$ for the $^{13}$C and the $^{12}$C layers in the 2-LG sample under study points to an equal charge accumulated on the two layers at each particular potential.[3] It should be noted that the bottom layer is not in contact with the electrolyte, whereas both layers are in an ohmic contact. The counterions from the electrolyte thus have to compensate the charge on the bottom layer through the top layer. The identical evolution of the doped bottom and top layers in 2-LG suggests only a weak influence of the electrolyte. In other words, solely the alterations caused by the electric field are manifested in the behavior of the bottom layer, while, apart from the electric field, the top layer is affected also by the double layer formed at the graphene/electrolyte interface. There is no indication of Li$^+$ intercalation in between the layers, not even at the lowest potential of -1.5 V.

The Raman intensities of both the G and G' bands in the 2-LG are being reduced as the potential magnitude is increasing for both doping directions, in a similar fashion for the top and the bottom layer.[3] There is no visible stacking order of the two graphene layers and therefore no or only minor coupling between the layers could be awaited, in contrast to AB-stacked bilayers (see below). Hence, the individual layers in the 2-LG should behave as they do in 1-LG under electrochemical doping. However, that is obviously not the case, especially as far as the G mode intensity is concerned, where substantial differences can be observed.[3,5] The unexpected drop of the G mode intensity in the 2-LG sample might be caused by changes in the electronic coupling due to coulomb repulsion of the two layers. As shown in recent works, the interaction between misoriented layers of graphene can result in the emergence of Van Hove singularities in the electronic states and/or leveling out of the electronic bands at certain angles of misorientation, in dependence on the excitation energy.[37-40] In analogy to the general case studied above, one can expect that the decoupling of the layers through doping can revert the electronic structure and thereby quench any enhancement effects caused by the interlayer interactions. Indeed, a gradual cancelation of the G band enhancement in bilayers twisted by the critical angle was observed *in-situ* during an electrostatic gating experiment.[41] On the other hand, no changes in the G band intensity were observed in bilayers twisted by an angle deviating far from the critical angle.

Variations in the 2D band (appearance or disappearance of an additional 2D$^+$ component) upon charging were observed in the same experiment.[41]

In case of 3-LG the combination of isotopically pure graphene layers is not sufficient to distinguish individual graphene layers. A mixing of isotopes is needed to produce a shift of the Raman bands to a suitable position between the peaks of pure $^{13}$C and $^{12}$C graphene.[4] The straightforward case is the 3-LG assembly composed of layers of natural $^{12}$C graphene on top, a 1:1 mixture of $^{12}$C:$^{13}$C in the middle, and pure $^{13}$C graphene at the bottom, see Figure 6. Combining these three, randomly stacked monolayers, 3-LG can be retrieved. The frequencies of the Raman modes in the stacked layers are found to be shifted with respect to those of 1-LG. The variations in frequency shifts may be linked to the variations in doping,[5] stress,[42] and interlayer interactions,[40] and these differences evidently vary for each specific layer as they do also from the three 1-LG layers on the substrate. Hence isotopically engineered few layer graphene is an ideal system to explore the influence of the environment and the substrate. In the 3-LG the bottom layer is in contact with the substrate, the top layer with the environment, and the middle one solely with graphene layers on either side.[4] In contrast, the 1-LG samples are in a direct contact with both the substrate and the environment.

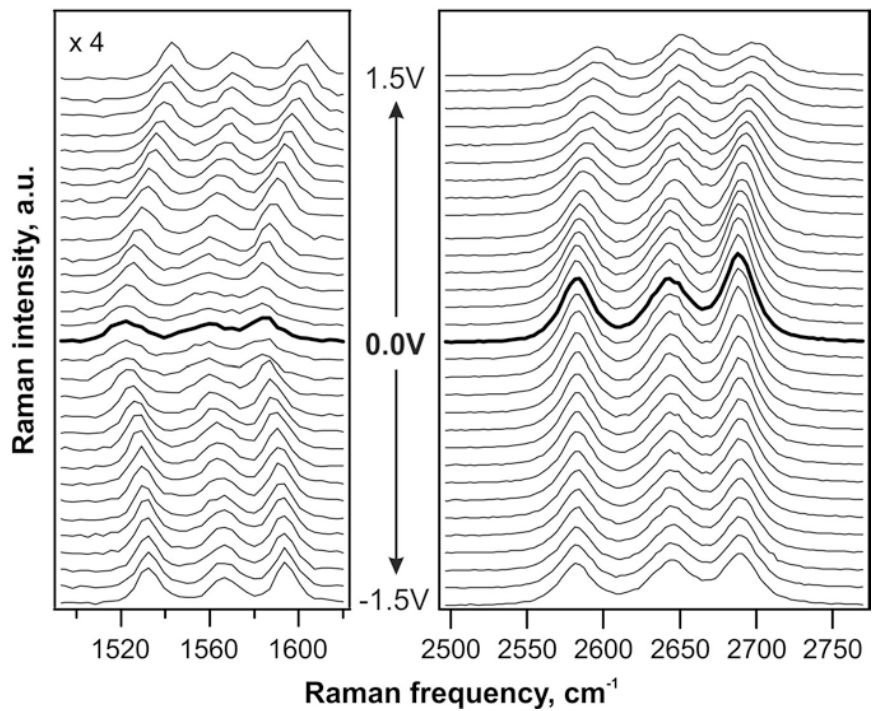

**Figure 6.** *In-situ* Raman spectroelectrochemical data for $\omega_G$ (left) and $\omega_{G'}$ (right) of a 3-LG sample, excited by 2.33 eV laser excitation energy. The bottom layer is $^{13}C$, the middle is $^{12/13}C$, and the top is $^{12}C$. Adapted with permission from Ref [4]. Copyright 2012 American Chemical Society.

Figure 6 shows an example of *in-situ* Raman spectroelectrochemical measurements taken on 3-LG.[4] Similarly, as in the case of 1-LG[5] and 2-LG,[3] the Raman features of 3-LG[4] show a notable dependency on the doping. The evolution of specific layers in 3-LG reflects their position in the stack. In the case of Ref [4], all 3 layers were contacted as the working electrode, but each layer had a particular environment: graphene and substrate for the bottom layer, graphene and graphene for the middle layer, and graphene and electrolyte for the top one. The spectroelectrochemical data show, therefore, the impact of the environment on the evolution of the Raman features during electrochemical doping. Additionally, the influence of the adjacency of electrolyte ions on graphene can be determined from the experimental data. The top layer only is in direct contact with electrolyte counterions, which compensate the charges injected into graphene layers. The middle and bottom layers are separated from the electrolyte by one and two layers of graphene, respectively.[4]

The electrode potential influences the Raman features of all 3 layers in the stack. Hence, neither the middle layer nor the bottom one are perfectly screened by the top layer. Fig. 7 shows the middle layer exhibiting an almost model evolution with only a slight asymmetry of the frequency shifts for the electron and hole doping.[4] On top of that, the $\omega_G$ fulfills the theoretical predictions at low doping levels:[28] there is a local maximum at 0 V and two local minima between ±0.1-0.2 V, which coincide with the energy of the G mode phonons. The largest asymmetry for the electron/hole doping is shown for $\omega_G$ of the bottom layer (Fig. 7), which illustrates a strong impact of the trapped charges in the $SiO_2$ support on the directly overlying graphene.[4]

Furthermore, there is no effect resembling the metallic screening of the electrostatic potential of the electrolyte ions in the case of graphene layers. This observation is consistent with the theoretical prediction of Guinea,[43] which proposed that the intra-layer hopping results in distribution of charges over many graphene layers.

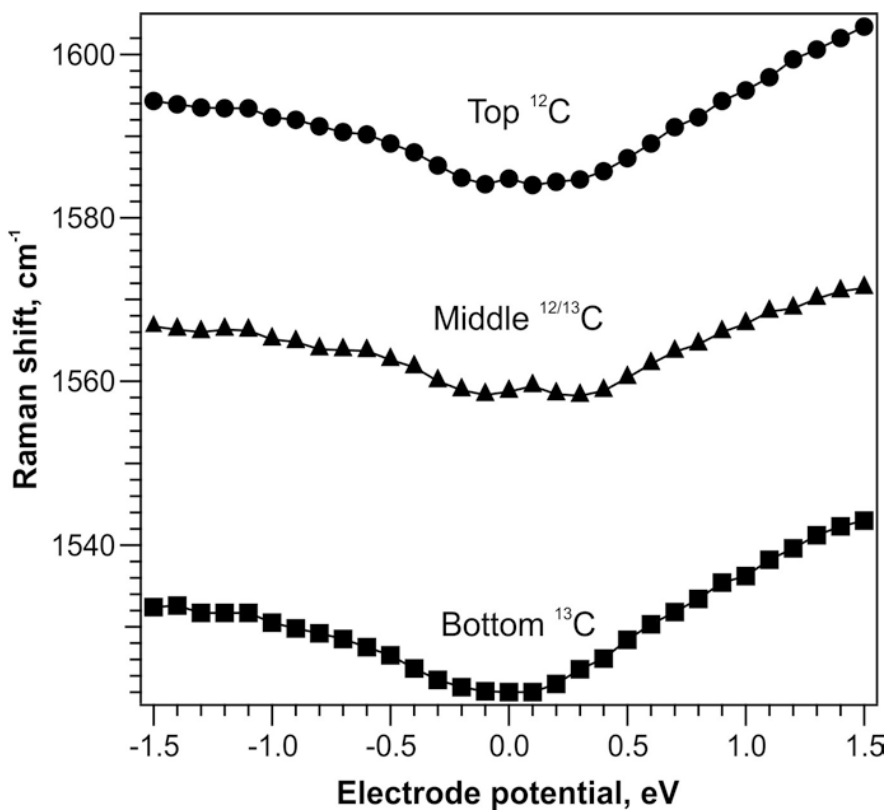

**Figure 7.** *In-situ* Raman spectroelectrochemistry of the G-band of a 3-LG sample, excited by the 2.33 eV laser excitation energy. Reprinted with permission from Ref [4]. Copyright 2012 American Chemical Society.

The results above describe the situation in randomly oriented graphene layers, where the turbostratic few-layer graphene can be easily obtained by the subsequent transfer of individual, isotopically labeled layers onto the target substrate.[3,4,27] However, the rotation angle between the graphene layers cannot be controlled very well and also the sequential transfer routine may introduce impurities between the layers. Hence, it is hardly possible to produce AB-2LG in this way. Recently, numerous works on graphene ad-layers (as-grown by the CVD procedure) appeared.[44-46] The ad-layers are, in general, multilayer seeds evolved during the deposition of a primary 1-LG using a copper catalyst. AB-stacked regions are often formed in the ad-layers. If the isotope composition of the CH$_4$ precursor and the growth conditions are properly controlled, the continuous layer can be composed of mostly one isotope (either $^{12}$C or $^{13}$C) and the ad-layer consists of the second isotope.[27] Moreover, regions with both AB and turbostratic configurations can be found within the same grains, which allows a comparison to be made between the response of the external perturbations regarding their dependence on the stacking order.[9] In such a specimen, the individual layers can be addressed by Raman spectroscopy, to monitor the impact of phonon self-energy renormalizations for each particular layer independently and to deepen the knowledge about the interlayer interactions. The AB 2-LG reveals two separate G modes, designated LG (lower frequency) and HG (higher frequency) due to a mass-related inversion symmetry-breaking.[9] These modes are associated with a symmetric (LG) and an anti-symmetric (HG) combination of E$_g$ and E$_u$ modes. Normally, only the E$_g$ is a Raman-active mode in the $^{12/12}$C AB-stacked 2-LG, where no mass-related symmetry-breaking occurs. In the case of $^{12/13}$C AB 2-LG, this symmetry is naturally lifted, because of the different isotopes constituting the unit cells in the top and the bottom layers. This is different from $^{12/13}$C turbostratic 2-LG, where the two separate G peaks are just linked to the E$_g$ modes from the non-interacting $^{13}$C and $^{12}$C layers. Electrochemical doping has different effects on the AB and turbostratic 2-LG (Figure 8).[3,9] In the case of the AB 2-LG, the frequency shifts of the LG and HG modes are smaller and more complex and also the intensity ratio between these two modes shows a distinct and characteristic evolution with the applied potential.[9] An explanation can be proposed by employing the band-gap opening as another process involved in the doping of the AB 2-LG. The frequencies of LG and HG modes decrease for negative potentials in the AB 2-LG, meaning that the variations in the C-C bonds are responsible for the observed effects in this potential region. In the AB 2-LG, the electrochemical doping reflects a lesser sensitivity of E$_F$ to the electrode potential because of the distinct electronic structure, in contrast to that for the turbostratic 2-LG. Additionally, spectroelectrochemical data from AB-2LG manifested less charge present on the bottom layer than on the top layer, a situation which is analogous to a device with a fixed potential at the bottom gate (realized through the permanent doping from the silicon substrate) and a variable potential at the top gate (realized electrochemically by varying the applied voltage).[9]

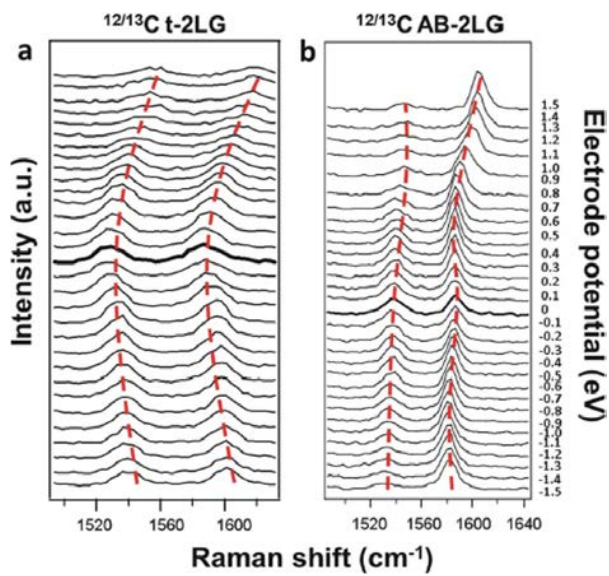

**Figure 8.** *In-situ* Raman spectroelectrochemistry of the G-modes in 12C/13C turbostratic (a) and AB-stacked (b) bilayer graphene, excited by the 2.33 eV laser excitation energy. Reprinted from Ref [9].

**Conclusions**

We have shown some of the possibilities of using *in-situ* Raman spectroelectrochemistry to study graphene generally and CVD graphene in particular. The use of electrochemical experiments allows researchers to reach higher doping levels together with a better control of the potential through the three-electrode setup, when compared to other methods for controlling the Fermi level. One of the most important consequences of the higher applied potentials is the observation of a cancellation of the interference effects causing a dramatic increase of the G band intensity at positive potentials. In particular, lowering the Fermi level down to one half of the excitation photon energy allows more detailed quantitative studies of the effects of electrons and hole doping of semiconductors. Furthermore, the employment of isotope labeling in the CVD process opens up other opportunities in addressing individual layers in multi-layered systems when carrying out Raman experiments. Electrochemical doping is also shown to allow studies of 2-LG graphene with distinct and different stacking orders, as well as in distinguishing spectroscopically the effects of charging the bottom, middle or top layers in a 3-LG system.

**Acknowledgements**

The authors (O.F. and M.K.) acknowledge the support of MSMT Project (LH13022), MSMT ERC-CZ Project (LL1301), and European Union FP7 Programme (No. 604391 Graphene Flagship), while M.S.D. acknowledges grant NSF-DMR 10-04147.

**Author biographies**

**Dr. Otakar Frank**

Otakar Frank is a junior group leader within the Department of Electrochemical Materials at J.Heyrovsky Institute of Physical Chemistry of the Academy of Sciences of the Czech Republic in Prague. He obtained his Ph.D. degree in 2005 at the Charles University in Prague. His main research interests include mechanical and electronic properties of graphene and related nanostructures.

**Prof. Mildred S. Dresselhaus**

Mildred Dresselhaus is an Institute Professor at MIT in the departments of Electrical Engineering and Physics, and has been active in materials research since the 1950s. Recent research activities in the Dresselhaus group that have attracted attention are in the areas of carbon nanotubes, bismuth nanostructures, and low-dimensional thermoelectricity.

**Dr. Martin Kalbac**

Martin Kalbac graduated in inorganic chemistry from the Charles University, Prague, Czech Republic (1998) where he also received a degree PhD. in 2002. Since 2001 he has worked at the J. Heyrovsky Institute of Physical Chemistry of the AS CR in Prague, as a research scien-tist. From 2010 he is a vice-director of the institute and from 2013 he leads the Department of Low Dimensional Systems. His research interest includes spectroscopy and spectroelectrochemistry of carbon nanotubes and graphene.

**Conspectus graphic**

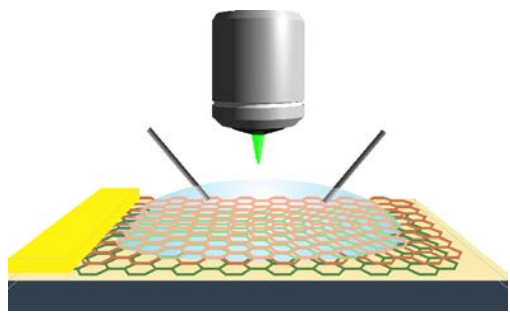